\documentclass[a4paper,fleqn]{cas-sc}

\def\tsc#1{\csdef{#1}{\textsc{\lowercase{#1}}\xspace}}
\tsc{WGM}
\tsc{QE}
\tsc{EP}
\tsc{PMS}
\tsc{BEC}
\tsc{DE}

\begin{document}
\let\WriteBookmarks\relax
\def\floatpagepagefraction{1}
\def\textpagefraction{.001}
\shorttitle{Rotating Magnetocaloric Effect in First-order Phase Transition Material Gd$_5$Si$_2$Ge$_2$}
\shortauthors{R. Almeida et~al.}

\title [mode = title]{Rotating Magnetocaloric Effect in First-order Phase Transition Material Gd$_5$Si$_2$Ge$_2$}

\author[1]{Rafael Almeida}
\credit{Conceptualization, Software, Formal analysis, Investigation, Data Curation, Writing – Original Draft, Writing – Review \& Editing, Visualization}

\author[2]{Rodrigo Kiefe}
\credit{Methodology, Software, Validation, Writing – Review \& Editing}

\author[1]{Ricardo Moura Costa Pinto}
\credit{Investigation, Writing – Review \& Editing}

\author[2]{João Sequeira Amaral}
\credit{Conceptualization, Investigation, Writing – Review \& Editing, Funding Acquisition}

\author[3,4]{Kyle Dixon-Anderson}
\credit{Investigation, Formal analysis, Resources, Writing – Review \& Editing}

\author[3,4]{Yaroslav Mudryk}
\credit{Resources, Writing – Review \& Editing, Supervision, Funding acquisition}

\author[1]{João Pedro Araújo}
\credit{Writing – Review \& Editing, Supervision, Funding acquisition}
                
\author[1]{João Horta Belo}
\cormark[1]
\credit{Conceptualization, Investigation, Writing – Review \& Editing, Supervision, Funding acquisition, Project administration}

\cortext[cor1]{Corresponding author: jbelo@fc.up.pt}

\affiliation[1]{organization={IFIMUP – Institute of Physics for Advanced Materials, Nanotechnology and Photonics},
                country={Department of Physics and Astronomy, Faculty of Sciences of University of Porto, Rua do Campo Alegre, 687, 4169-007 Porto, Portugal}}
\affiliation[2]{organization={Physics Department and CICECO — Aveiro Institute of Materials},
                country={University of Aveiro, 3810-193 Aveiro, Portugal}}
\affiliation[3]{organization={Division of Materials Science and Engineering, Ames Laboratory of US DOE},
                country={Iowa State University, Ames, Iowa 50011-3020, United States}}
\affiliation[4]{organization={Department of Materials Science and Engineering},
                country={Iowa State University, Ames, Iowa 50011-2300, United States}}

\begin{abstract}
The rotating magnetocaloric effect (RMCE) induced by self-demagnetization has been investigated in the giant magnetocaloric effect (GMCE) material Gd$_5$Si$_2$Ge$_2$. This shape-dependent effect had thus far only been reported in pure Gd, marking this as the first analysis of the effect in a sample with a magnetostructural first-order phase transition. By rotating the applied magnetic field vector while keeping its intensity constant, the demagnetizing field within a high-aspect ratio sample changes significantly, resulting in a RMCE. We characterize RMCE by determining the adiabatic temperature change ($\Delta T_{ad}^{rot}$) directly through temperature measurements, and the isothermal entropy change ($\Delta S_M^{rot}$) via magnetometry and magnetostatic simulations. We obtain a remarkable maximum $\Delta T_{ad}^{rot}$ of 1.77 K for a constant external field of 0.8 T, higher than that obtained under 1.0 T. The magnetostatic simulations not only corroborate the highly non-monotonous field-dependence of $|\Delta S_{M}^{rot}|$, which reaches 95\% of its maximum value at 0.8 T, 6.12 J K$^{-1}$ kg$^{-1}$ for the experimentally measured shape, but also estimate a 35\% increase in the maximum $|\Delta S_{M}^{rot}|$ up to 8.67 J K$^{-1}$ kg$^{-1}$ in a simulated shape with higher aspect ratio.
\end{abstract}


\maketitle

\section{Introduction}
\label{sec:Intro}

The magnetocaloric effect (MCE) has been intensely investigated in the past three decades due to its potential application in highly efficient heat pumps without vapor refrigerants \cite{Tishin2016, Kitanovski2015,Franco2018,IIR2022Caloric,Klinar2024}.

Although the MCE has been known as a basic property of magnetic materials for over 100 years \cite{Thomson1878,Smith2013}, it was not considered for room temperature refrigeration until 1976, when G. Brown reported a device capable of generating a 47 K temperature span in a water-ethanol mixture with gadolinium \cite{Brown1976,Smith2012}. Two years later, in a report that went unpublished until 2014 \cite{Brown2014regenerationtestsroomtemperature}, G. Brown and S. S. Papell showed that the same device was capable of generating an $\mathrm{80\: K}$ temperature span by increasing the amount of Gd used and improving other components to decrease fluid mixture.

The interest in the MCE was renewed around three decades ago with the discovery of alloys displaying magnetostructural first-order phase transitions (FOPT) that lead to a giant magnetocaloric effect (GMCE) at room temperature, nearly double the peak intensity of the MCE in Gd, which displays a magnetic second-order phase transition (SOPT) \cite{NIKITIN1990363,Pecharsky1997}. Since then, several other prominent material families with GMCE near ambient temperature have been discovered and optimized, such as La-Fe-Si, Mn-Fe-P-Si, Ni-Mn-In, and others \cite{Liu2019,LIONTE202143,LAI2023167336,Law2023}.

This study focuses on one of the first-order phase transition materials exhibiting a GMCE - Gd$_5$Si$_2$Ge$_2$. It is part of the larger family of alloys, Gd$_5$(Si$_x$Ge$_{1-x}$)$_4$, with varying properties and phase transition temperatures between 40 K (x=0) and 340 K (x=1) \cite{Pecharsky1997,MELIKHOV2015143}.

With few exceptions, RMCE has been mainly studied in single-crystals or textured polycrystals with magneto-crystalline anisotropy (MCA) \cite{Nikitin2010,Balli2014,Zhang2015}. Despite the wide recognition of the relevance of the demagnetizing effect to the MCE \cite{Bahl2009,Kuzmin2011,Nielsen2012,Trevizoli2012,Romero-Muniz2014}, few studies demonstrate its potential to generate a RMCE by rotating a high aspect-ratio sample within a constant magnetic field. The first to accomplish this was a patent published by J. Barclay et al. in 1984, where an adiabatic temperature change ($\Delta T_{ad}$) induced by external field rotation ($\Delta T_{ad}^{rot}$) is presented in a Gd sample under a constant magnetic field of 0.3 T. Recently, a 2023 study featured an analogous experiment for a Gd sample rotating under a higher magnetic field ($\mathrm{1.6\: T}$) and finally, in early 2024, a detailed magnetic field-dependent study revealed the highly non-monotonous field-dependence of the demagnetizing field-induced RMCE in Gd \cite{Barclay1984,Badosa2023,Almeida2024}. 

One of the main challenges for mass adoption of magnetic refrigeration systems is the cost of generating magnetic field. The permanent magnet structures used are often the most expensive part of the devices \cite{RUSSEK20061366}, and the work required to alternate the magnetic field value on the magnetocaloric material to induce an MCE can represent over 50\% of the total input power \cite{TURA2011628}. To modulate the magnetic field intensity, conventional magnetic refrigeration systems cyclically shift parts of the permanent magnet structures, or move the magnetocaloric material from high to low magnetic field zones \cite{BJORK20103664,BJORK2010437}. On the other hand, a system based on the RMCE only requires a change in the magnetic field orientation, so the magnetocaloric material may remain inside a high magnetic field zone throughout the thermodynamic process. Furthermore, the above-mentioned magnetic field-dependent study revealed that the RMCE of Gd was most competitive with the conventional MCE when using low magnetic field intensities, which is also a way of significantly reducing the amount of permanent magnet material required \cite{Almeida2024}. This suggests that RMCE could be a viable alternative basis for magnetic refrigeration devices that aim to use low magnetic field intensities.

While the rich physics of Gd$_5$Si$_2$Ge$_2$ have already been extensively studied since its discovery in 1997 \cite{Pecharsky2001,Pecharsky2003,Mudryk2005}, the current work is the first to measure and extensively characterize RMCE induced by the demagnetizing effect in this or any other FOPT GMCE material.

\section{Methods and experimental details}
\label{sec:Methods}

We characterize RMCE of Gd$_5$Si$_2$Ge$_2$ as a function of sample temperature and external field amplitude in two complementary approaches: direct measurements of temperature to determine $\Delta T_{ad}$ and $\Delta T_{ad}^{rot}$, and magnetometry to determine the isothermal entropy change on field application  ($\Delta S_{M}$). Magnetometry is then complemented with simulations from the 3D finite element magnetostatics simulation software \textit{FEMCE} (Finite Element Magnetocaloric Effect) \cite{Kiefe2025} to enable the determination of the isothermal entropy change on field rotation $\Delta S_{M}^{rot}$. The direct measurements were done with an applied magnetic field up to 1.0 T, while the magnetometry study was done up to $\mathrm{2.0\:T}$.

\subsection{Sample preparation}

A 6 g button of Gd$_5$Si$_2$Ge$_2$ was prepared through arc-melting, using high-purity gadolinium metal from the Materials Preparation Center of the Ames National Laboratory. Ge and Si were commercially sourced. The weight losses during melting were less than 1 wt.\%. The sample was annealed to improve homogeneity. Magnetic and structural characterization confirmed the intended phase was manufactured and can be found in appendix A.

\begin{figure}
    \centering
    \includegraphics[width=140 mm]{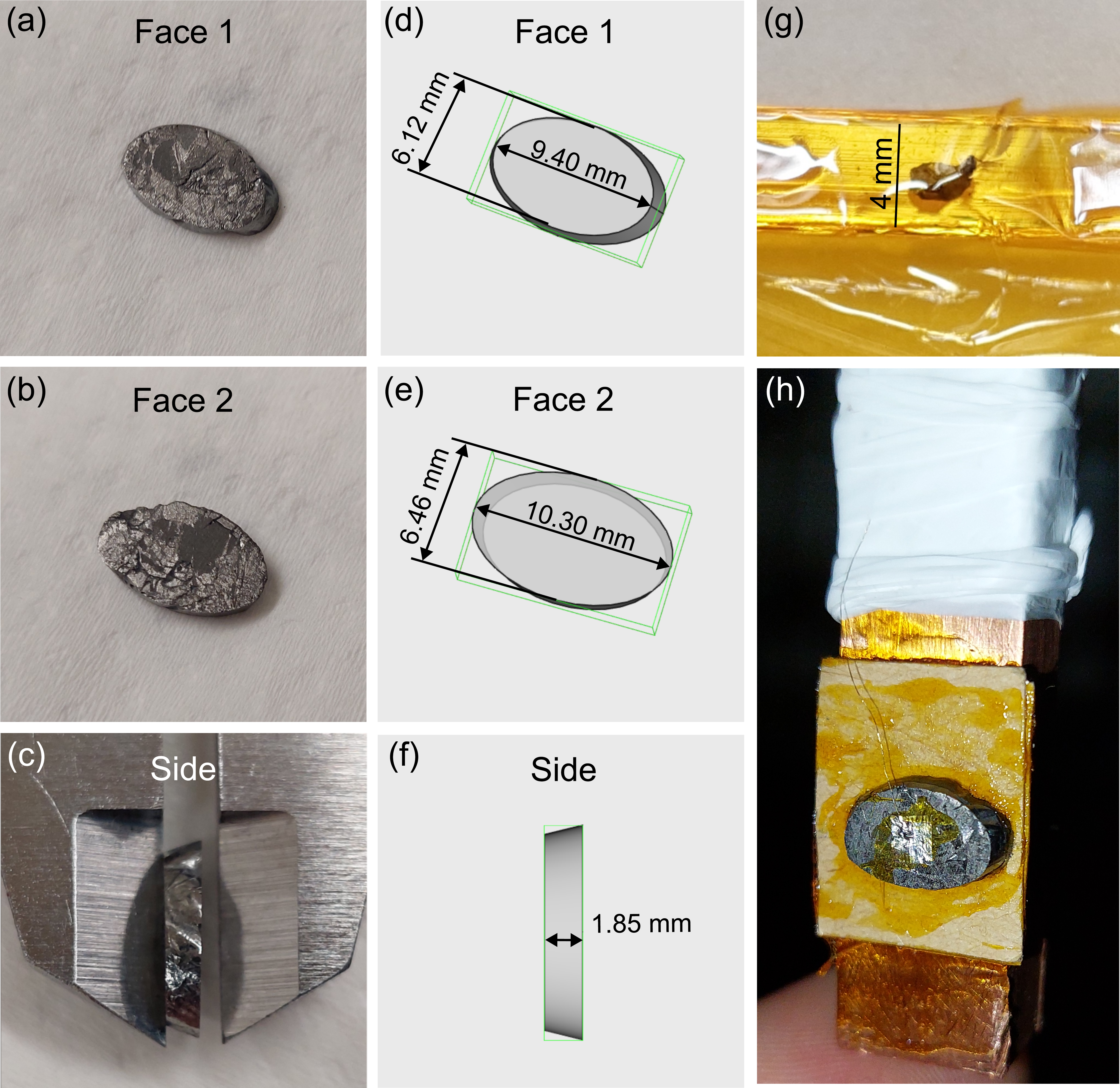}
    \caption{Photos of the sample used for direct temperature measurements (a-c) and corresponding pictures of the CAD 3D model used for the magnetostatic simulations (d-f). (g) The smaller sample measured in the MPMS, mounted on a quartz sample holder with polyimide tape. (h) The sample mounted on the cryostat, with thermocouple installed.}
    \label{fig:1}
\end{figure}

For the study of the demagnetizing field-induced RMCE, a high aspect-ratio shape is required so that the demagnetizing field is substantially different when the applied field is oriented along different directions, and a significant change in internal field may be induced by rotation \cite{Almeida2024}. To this end, the approximately ellipsoid shape of the cast button of the sample was cut in half, such that the region with the widest cross-section was accessed. Then, the thickness was reduced to 1.8 mm. For maximizing the demagnetizing field-based RMCE, a narrower sample would be beneficial (this is explored further ahead), but the brittle nature of the alloy would make it susceptible to breaking during shaping or measurements. The final sample weighed 628.3 mg, and its shape can be seen in figure \ref{fig:1}. 

\subsection{Direct temperature measurements}

To measure the sample temperature upon magnetic field changes, a type-K thermocouple with 25 $\mu$m-diameter wire was used. Typically, the thermocouple sensing tip is sandwiched between two halves of the sample, allowing a remarkably fast time-response \cite{Salazar2023}, but this would decrease its overall aspect ratio. Instead, a similar effect was achieved while keeping the sample thickness as low as possible by using a negligible amount (less than 1 mg) of aluminium foil to press the thermocouple tip against the sample, removing excess adhesive (GE-varnish), and fully surrounding the thermocouple tip, improving thermal contact without compromising the demagnetizing field-based RMCE (see figure \ref{fig:1}(h)).

The sample was mounted in a closed-circuit cryostat under vacuum (P$<$10$^{-4}$ mbar). Two thin layers of craft paper acted as a thermal resistance to slow down the heat exchange between the sample and the copper base, further improving adiabatic conditions. The field source is an electromagnet for field intensities between 0.2 T and 0.8 T, or a Halbach array of permanent magnets for 1.0 T. For field applications/removals, the magnetic field source is physically moved such that the sample is inserted and removed from the magnetized region in less than 1 second. For the field rotations, the sample is constantly in the magnetized region. The Halbach array is rotated in less than 1 second, and the electromagnet in about 2 seconds. These time intervals are only about 4\% of the thermal exchange time-constant ($ \mathrm{45\:s}$), such that an exponentially decaying temperature would only be underestimated by 5\% (see figure \ref{fig:2}). Temperature acquisition and control was done with a \textit{Lakeshore 336} temperature controller. 

The thermocouple's signal noise level is 0.02 K, making data processing the determining factor for uncertainty. Namely, how much time is given for thermal equilibrium to occur after a field change. Choosing between reasonable variations of that time (2$\pm$1 s) results in a maximum variation of the obtained $\Delta T_{ad}$ below $\pm$0.1 K, which is below 10\% for most of the presented measurements.

Details on the measurement protocols are given in the results section.

\subsection{Magnetometry}

A smaller, irregularly shaped fragment with $m=6.93$ mg picked from the same cast button (see figure \ref{fig:1}(g)) was used for the magnetization measurements. The \textit{MPMS3} commercial SQUID magnetometer was used to measure magnetization versus temperature and magnetic field. The absolute values measured were corrected following the geometry-independent procedure described in reference \cite{Amorim2021}. 

\subsection{Isothermal entropy change calculation}

The determination of $\Delta S_M$ from magnetization measurements is done through the numerical integration of the Maxwell relation 

\begin{equation}
\label{eq:1}
    \Delta S_M(T,H_f,H_i) = \mu_0 \int _ {H_i}^{H_f} \frac{\partial M(T,H)}{\partial T}dH.
\end{equation}

However, the accuracy of this approach for materials exhibiting a first-order phase transition has been questioned in the past due to two main issues: the non-homogeneity of samples due to phase coexistence, and the non-equilibrium states exhibited by materials with metastability \cite{Amaral2010}.

Indeed, Gd$_5$Si$_2$Ge$_2$ displays both of these phenomena. In this work, we used the following approaches that have been found to successfully mitigate the errors produced by these phenomena:
\begin{itemize}
    \item measured a small sample (6.93 mg), to facilitate thermal equilibrium with the instrument;
    \item performed isofield measurements under cooling with a slow temperature sweep rate of 2 K/min, instead of isothermal measurements;
    \item raised the temperature well above $T_C$ (350 K) between each measurement;
    \item used a small temperature step to avoid ``smearing'' the $\Delta S_M$ curve and small field step to avoid introducing artificial oscillations from lack of points;
    \item corrected for demagnetization by measuring an isothermal magnetization at a temperature well below $T_C$ (50 K).
\end{itemize}

The necessity and validity of these approaches has been discussed in detail in previous works \cite{Amaral2010,Das2010,Caron2009,Bez2018}.  

While we use isofield measurements for the analysis, we also acquired two datasets of magnetization in isothermal conditions  (increasing
field followed by decreasing field), which help demonstrate the concepts detailed in this section. The isotherms were measured with a magnetic field change rate gradually increasing from 0.5 mT/s (low fields) to 3.5 mT/s (high fields). At every temperature, each pair of isothermal measurements is preceded by a thermal reset in which the sample temperature is increased to 350 K. 

\subsection{Magnetostatic simulations}

The demagnetizing field for samples of highly symmetric shapes, such as rectangular prisms, may be computed analytically by assuming homogeneous magnetization \cite{Aharoni1998}. However, the demagnetizing field and the magnetization, which are interdependent, are often non-homogeneous. Furthermore, an analytical approach is unfeasible for most sample shapes or assemblies of several magnetic bodies. This makes the numerical computation of the demagnetizing field the best approach for accurate determination of the demagnetizing field. This has been previously achieved for assemblies of rectangular prisms by discretizing the shapes into a high number of homogeneously magnetized rectangular cells, combined with the analytical result for homogenously magnetized rectangular volumes \cite{Smith2010,Christensen2011}. More recently, the open access tool \textit{FEMCE} employed a finite element method to solve the Maxwell magnetostatic equations for arbitrary shapes \cite{Kiefe2025}. 

In this study, we use \textit{FEMCE} to obtain the 3D magnetic field distribution within the sample shape at each temperature, applied magnetic intensity, and orientation. To perform these simulations, the workflow was the following: 1) measure magnetization vs. temperature and magnetic field as detailed in the previous section; 2) correct the data for demagnetization, removing existing geometric effects from the dataset (to this end, a magnetization isotherm measured at 50 K was used); 3) export the magnetization data in matrix format, as well as the corresponding temperatures and magnetic fields, to import in \textit{FEMCE}; 4) import the 3D CAD design of the sample shape; 5) create the finite elements mesh for of simulation volume; 6) choose the temperature, applied magnetic field intensity, and orientation. To perform the simulations in sets, we use a version of \textit{FEMCE} without GUI.

\begin{figure}
    \centering
    \includegraphics[width=70 mm]{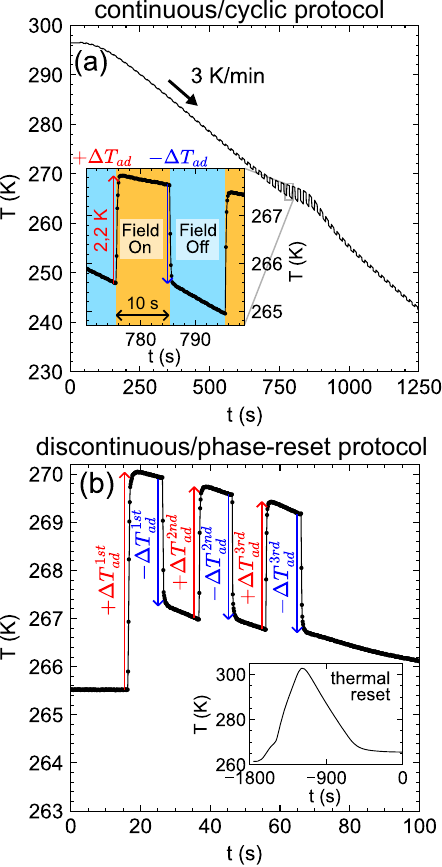}
    \caption{The two direct temperature measurement protocols. (a) Continuous/cyclic protocol under cooling, in which the heat bath temperature of the sample is continuously changing and the field is applied and removed every 10 seconds. (b) The discontinuous/phase-reset protocol, in which the heat bath temperature of the sample is stabilized prior to the field being applied and removed (also every 10 seconds). The discontinuous protocol measurements are always preceded by a thermal reset taking around 30 minutes, as shown in the inset.}
    \label{fig:2}
\end{figure}

\section{Results and discussion}
\label{sec:Results}

We start with the discussion of the direct measurements of the more familiar conventional MCE, followed by the direct measurements of RMCE. We follow the same logic for the indirect measurements (magnetic entropy change), ending with a discussion of the RMCE exhibited for a higher aspect ratio sample. 

\subsection{Direct temperature measurements of the conventional magnetocaloric effect}

Measurements of $\Delta T_{ad}$ at different temperatures in samples without hysteresis can be done by cycling the magnetic field while slowly ramping the heat bath temperature (heating or cooling). This has been referred to as a continuous \cite{Skokov2012,CHIRKOVA201615}, or cyclic protocol \cite{Cugini2020}. Within this protocol, we have alternated the magnetic field every 10 seconds, which is exemplified in figure \ref{fig:2}. For samples with significant thermal hysteresis, each set of measurements is preceded by a thermal reset, where the sample is heated well above the phase transition temperature, as also done in the magnetization measurements. In this case, the $\Delta T_{ad}$ obtained after the first magnetic field application ($+\Delta T_{ad}^{1st}$) will be clearly different from what is obtained after subsequent applications ($+\Delta T_{ad}^{2nd}$ and $+\Delta T_{ad}^{3rd}$), as shown in figure \ref{fig:2}(b). This is referred to as the discontinuous \cite{Skokov2012,CHIRKOVA201615}, or phase-reset \cite{Cugini2020} protocol.

\begin{figure}
    \centering
    \includegraphics[width=70 mm]{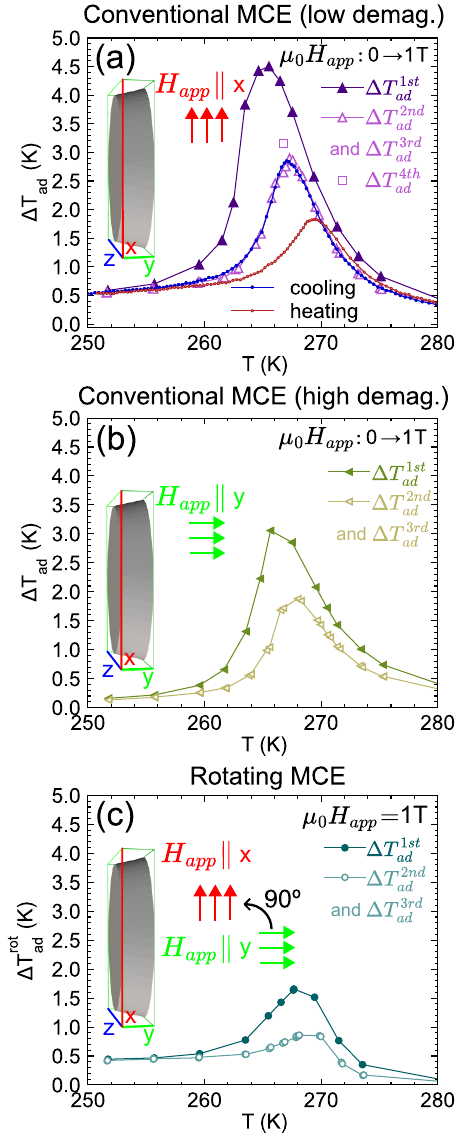}
    \caption{The directly measured $\Delta T_{ad}$ values obtained for (a) the low demagnetizing field orientation ($H_{app}\parallel y$) in different measurement protocols, and (b) the high demagnetizing field orientation ($H_{app}\parallel x$). (c) The $\Delta T_{ad}^{rot}$ obtained by changing the magnetic field orientation from $H_{app}\parallel x$ to $H_{app}\parallel y$. The "cooling" and "heating" curves in (a) refer to the continuous protocol.}
    \label{fig:3}
\end{figure}

In figure \ref{fig:3}, we show the adiabatic temperature change of the conventional MCE (field application) in (a) a low demagnetizing field orientation (parallel to the sample's larger axis, x, labelled $H_{app}\parallel x$), (b) a high demagnetizing field orientation (parallel to the sample's shortest axis, y, labelled $H_{app}\parallel y$), and (c) the RMCE obtained by rotating the magnetic field from the high demagnetizing field orientation to the low demagnetizing field orientation ($H_{app}\parallel y \rightarrow H_{app}\parallel x$). 

We obtained a remarkable value of maximum $+\Delta T_{ad}$=4.50 K when applying a 1.0 T magnetic field in the low demagnetizing field orientation at $\mathrm{265.5\:K}$. This is a noteworthy value, because few studies ever directly obtained $|\Delta T_{ad}|$ values above 4 K for such a low magnetic field \cite{Franco2018}. Two notable exceptions are Fe$_{49}$Rh$_{51}$, with -4.58 K, and Ni$_{45.7}$Mn$_{36.6}$In$_{13.5}$Co$_{4.2}$, with -4.35 K, both with inverse MCE measured for $\mathrm{1.0\: T}$ magnetic field applications following discontinuous protocols \cite{CHIRKOVA201615,Gottschall2015}. Regarding previous studies of Gd$_5$Si$_2$Ge$_2$, we are not aware of direct measurements of $\Delta T_{ad}$ for a 1.0 T magnetic field. Zeng et al. directly measured a maximum $\Delta T_{ad}$=3.9 K in Gd$_5$Si$_2$Ge$_2$ for a 1.5 T magnetic field, but they do not specify the measurement protocol or shape of the sample (which dictates the demagnetizing effect), making comparison difficult \cite{Zeng2012}. 

Figure \ref{fig:3}(a) also shows the $+\Delta T_{ad}$ values obtained in the second and third magnetic field applications, as well as the values obtained in the continuous protocol (labelled ``cooling'' and ``heating'') with the background temperature being changed at 1 K/min. The reversible $\Delta T_{ad}$ obtained in the 2nd and 3rd subsequent field applications is still very significant, reaching a maximum value of $\mathrm{3.16\: K}$ at a slightly higher temperature of 266.8 K. This underscores that Gd$_5$Si$_2$Ge$_2$ is one of the best magnetocaloric materials with highest $\Delta T_{ad}$ at 1.0 T even when considering cyclic conditions, as few materials cross the 3 K threshold \cite{Gottschall2019}.

The values of $+\Delta T_{ad}^{2nd}$ and $+\Delta T_{ad}^{3rd}$ measured in the discontinuous protocol have a remarkable agreement with those measured in the continuous protocol during cooling. Thus, the background temperature is varying slowly enough in the continuous protocol (1 K/min) to ensure nearly identical thermodynamic conditions as in the discontinuous protocol (the field applications are also done after cooling in the thermal reset). On the other hand, the curve obtained from the field applications during heating is clearly different: lower in amplitude and shifted to higher temperatures. This difference can be interpreted by the fact that during heating, a field application works against the temperature change in terms of direction of the phase transformation, hindering the MCE more significantly than during cooling, whereas the field applications force the phase transformation to proceed in the same direction as the background temperature change, achieving a partial thermal reset between each field change. Finally, we call attention to a measurement performed at a single temperature, in which the magnetic field is applied a fourth time (labelled $\Delta T_{ad}^{4th}$) within the discontinuous protocol after waiting for the temperature to fully reach the initial one. While $\Delta T_{ad}^{4th}$ is still clearly lower than $\Delta T_{ad}^{1st}$, it is a clear outlier from the $+\Delta T_{ad}^{2nd}$ and $+\Delta T_{ad}^{3rd}$ curve, implying that there is a range of values that the $+\Delta T_{ad}$ of subsequent field applications can take, depending on how long one waits for the sample temperature to return to the initial value.

The maximum $+\Delta T_{ad}$ obtained when applying a 1.0 T field in the high demagnetizing field orientation was 3.06 K, at 265.7 K. While still very significant, it is only 68\% of the maximum value obtained when the field was applied in the low demagnetizing field orientation. This highlights the importance of either minimizing or taking demagnetization into account when measuring the MCE. At the same time, the 32\% difference between the peak $+\Delta T_{ad}$ obtained for the different orientations already suggests the potential of achieving a high $+\Delta T_{ad}^{rot}$ exclusively through a change in the demagnetizing field. Indeed, in a material without hysteresis and other sources of anisotropy, it is expected that the demagnetizing field-based rotating MCE is approximately equal to the difference of the conventional MCE exhibited in each orientation \cite{Almeida2024}.

\subsection{Direct temperature measurements of the rotating magnetocaloric effect}

During the thermal reset performed for the RMCE measurements, the magnetic field remains applied along the high demagnetizing field orientation until the measurement temperature stabilizes. Then, the field is rotated every 10 s as in the discontinuous protocol, similarly to the measurements of the conventional MCE shown in figure \ref{fig:2}(b). 

Figure \ref{fig:3}(c) shows the $\Delta T_{ad}^{rot}$ (RMCE) obtained in the discontinuous protocol when rotating the applied magnetic field from the high demagnetizing field orientation to the low demagnetizing field orientation. The peak value of $\Delta T_{ad}^{rot}$ measured for 1.0 T was 1.66 K at 267.7 K, 37\% of the maximum value obtained for the conventional MCE in the low demagnetizing field orientation. Notably, the peak value of the RMCE closely corresponds to the difference of the conventional MCE amplitudes for the different field orientations (1.44 K). 

Some insight into the shape of the $\Delta T_{ad}^{rot}(T)$ curve seen in figure \ref{fig:3}(c) can also be given by examining the two $\Delta T_{ad}(T)$ curves shown in figures \ref{fig:3}(a) and (b): the starkest difference between these two curves occurs below the phase transition temperature at 1.0 T, around 270 K, since the magnetization increases drastically, in turn making the demagnetizing effect more significant. Consequently, the temperature profile of $\Delta T_{ad}^{rot}(T)$ is asymmetrical, having a higher tail towards low temperatures, where the internal magnetic field change between the two orientations is larger. This effect will be examined and explained through the magnetostatic simulations shown further ahead.

\begin{figure}
    \centering
    \includegraphics[width=140 mm]{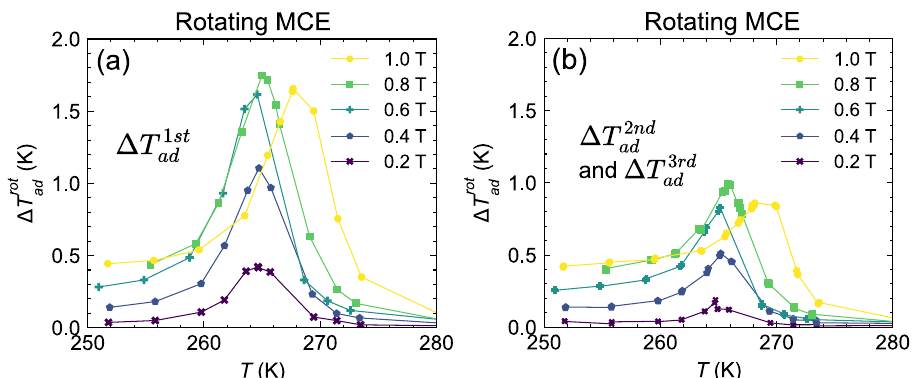}
    \caption{The directly measured $\Delta T_{ad}^{rot}$ obtained through the discontinuous protocol for the rotation of different applied magnetic field intensities on the (a) first magnetic field application (irreversible), and (b) subsequent magnetic field rotations. All values correspond to rotation from high to low demagnetizing field orientations, as illustrated in figure \ref{fig:3}.}
    \label{fig:4}
\end{figure}

Figure \ref{fig:4} shows the $\Delta T_{ad}^{rot}$ obtained through the discontinuous protocol for different applied magnetic field amplitudes. Considering the peak amplitudes obtained at different field intensities, a non-monotonous behaviour is observed, in which the peak amplitude very quickly increases, until 0.8 T, where it reaches its peak, and then decreases for the 1.0 T field amplitude. Simultaneously, the 0.8 T and 1.0 T curves show a slight shift to higher temperatures, which is distinct behavior than previously observed for the RMCE in Gd \cite{Almeida2024}. On the other hand, the non-monotonous behavior is quite comparable to what was found in Gd.

The highest $\Delta T_{ad}^{rot}$ measured overall was 1.77 K, obtained for the first 0.8 T field rotation at 265 K (figure \ref{fig:4}(a)). It is interesting to note that RMCE under the applied field as low as 0.6 T is already comparable to the RMCE obtained for the 1 T field. This emphasizes that any future prototype utilizing RMCE should aim at using low magnetic fields, as previously suggested \cite{Almeida2024}. Figure \ref{fig:4} (b) shows the $\Delta T_{ad}^{rot}$ values obtained for subsequent magnetic field rotations, which are all similarly attenuated due to the metastability of the sample, as shown for the measurements obtained in 1 T shown in figure \ref{fig:3}.

Furthermore, to our knowledge, this is the highest $\Delta T_{ad}^{rot}$ ever measured under a constant external magnetic field in a material without any MCA. Even for those with MCA, few materials reach values above 2 K for low fields, with some prominent examples being NdCo$_5$ for 2 T or NdCo$_{4.2}$Fe$_{0.8}$ for 1 T or 2 T \cite{WANG2020106676}. Such values may be achievable in Gd$_5$Si$_2$Ge$_2$ exclusively through the demagnetizing effect by shaping the sample in a shape with a higher aspect ratio, as suggested by the results of simulations shown in the following sections.

\subsection{Indirect measurements of the conventional magnetocaloric effect - magnetic entropy change during field application}

\begin{figure}
    \centering
    \includegraphics[width=140 mm]{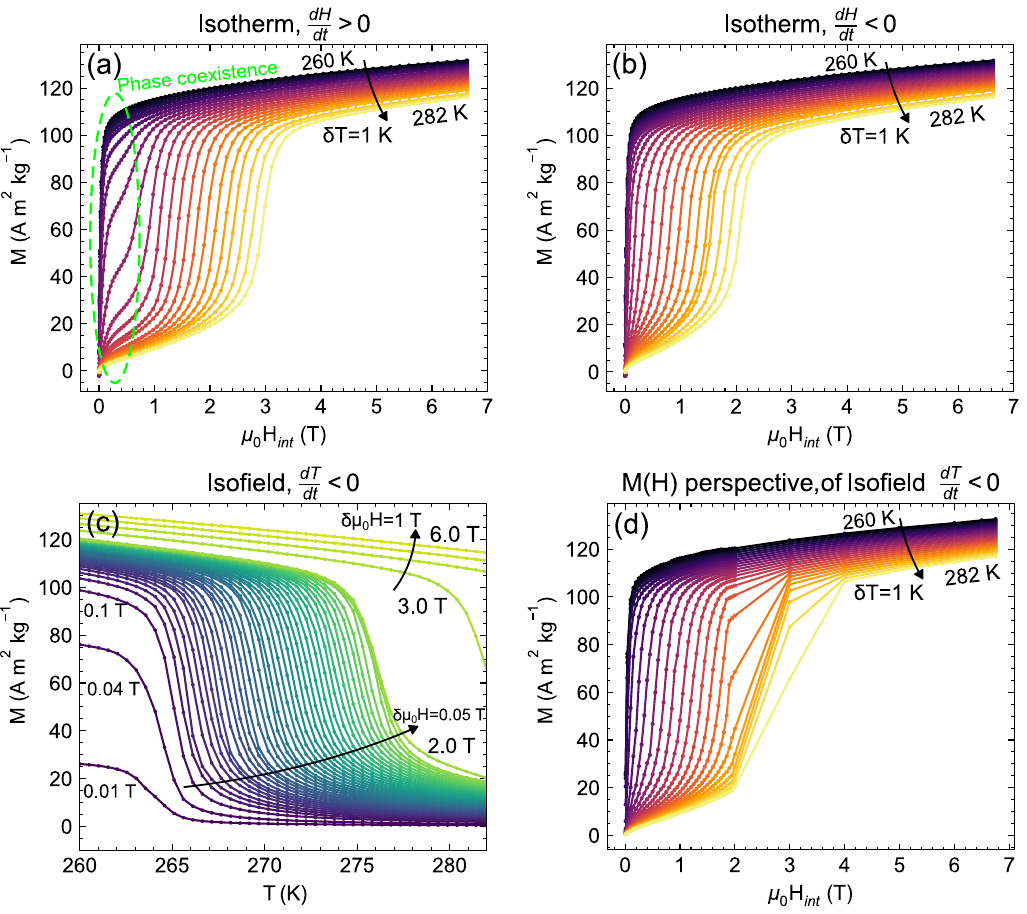}
    \caption{The different magnetization datasets acquired through different protocols. Magnetization isotherms measured with (a) increasing magnetic field and (b) decreasing magnetic field. Magnetization isofield curves measured under cooling with a constant applied magnetic field are plotted (c) versus temperature and (d) versus magnetic field, to facilitate comparison with the curves measured with constant temperature. The sample temperature is raised to 350 K between each pair of isotherms and each isofield measurement.}
   \label{fig:5}
\end{figure}

Figure \ref{fig:5} shows magnetization data sets acquired using different protocols mentioned in section \ref{sec:Methods}. Isothermal curves plotted from the isofield measurements performed as a function of temperature (figure \ref{fig:5}(d)) highly resemble the values acquired through isothermal measurements in decreasing field (figure \ref{fig:5}(b)). Contrastingly, the magnetization obtained in isothermal measurements with increasing field (figure \ref{fig:5}(a)) displays evidence of significant phase coexistence, even though the phase transformation direction is the same as during decreasing temperature. This makes the increasing field dataset prone to artificial peaks of $\Delta S_M$, as discussed in \cite{Amaral2010,Das2010}. The chosen dataset for our study were the isofield curves obtained under cooling shown in figure \ref{fig:5}(c) and figure \ref{fig:5}(d), as they more closely correspond to the thermodynamic path followed in the stable protocol of the direct temperature measurements, as these were also preceded by a thermal reset.

To determine $\Delta S_M$ taking demagnetization into account, we must first compute what the internal field is for each applied magnetic field intensity. This is where the magnetostatic simulations become necessary. These simulations allow us to estimate the internal magnetic field as a function of temperature and applied (external) field amplitude at each point of the sample's volume. In each computation, constant temperature throughout the sample is assumed. As such, the finite elements of the sample have the same M(H) response, corresponding to the isothermal magnetization at the simulation temperature. However, the converged solution of the system allows for a different internal magnetic field at each element.

\begin{figure}
    \centering
    \includegraphics[width=160 mm]{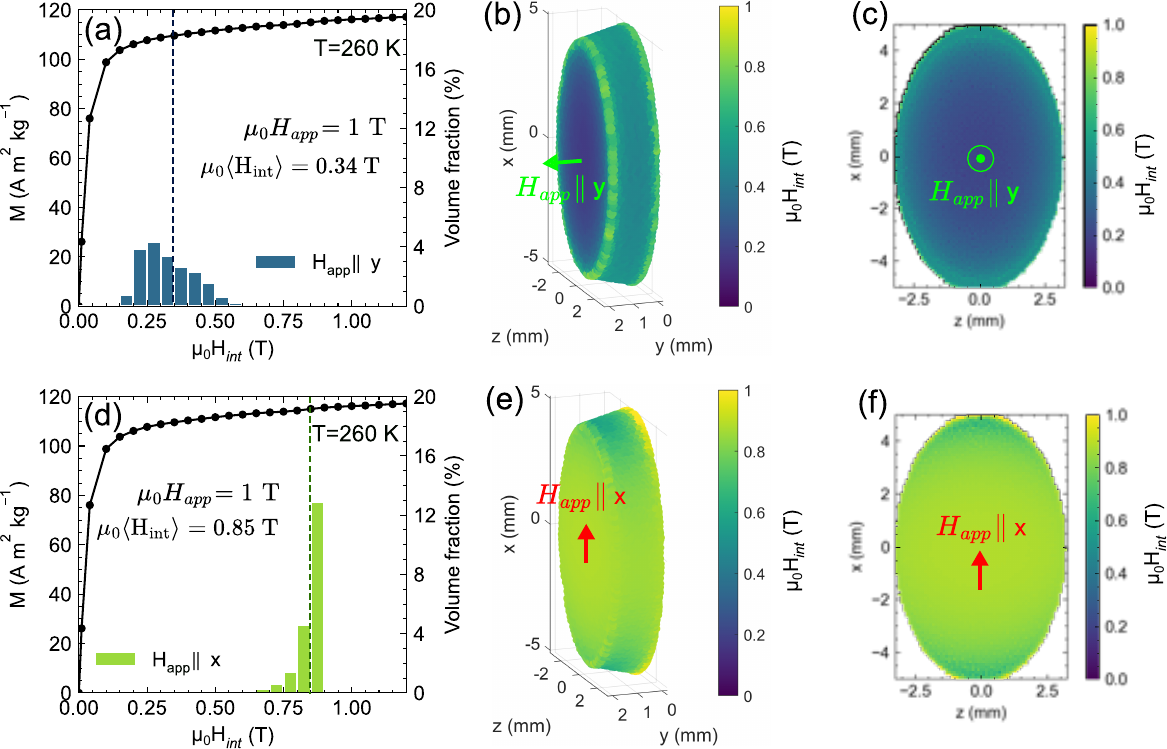}
    \caption{\textit{FEMCE} magnetostatic simulations for two magnetic field orientations at 260 K. The magnetization isotherm for the simulation temperature, shown together with the histogram of the internal magnetic field intensities for the (a) high demagnetizing field orientation and (d) the low demagnetizing field orientation. The results of the simulation in a 3D scatter plot for the (b) high and (e) low demagnetizing field orientations. The volume-average of the field intensities over the thickness of the sample is shown in 2D for the (c) high and (f) low demagnetizing field orientations.}
    \label{fig:6}
\end{figure}

Figure \ref{fig:6} shows the results of a pair of simulations done for 260 K and an applied field of 1.0 T applied for the high demagnetizing field ($H_{app}\parallel y$) and low demagnetizing field ($H_{app}\parallel x$) orientations. The sample was discretized into 41594 elements. The statistical distribution of internal fields for each orientation is shown in the histograms in figure \ref{fig:6}(a) and \ref{fig:6}(d). In this low temperature example, when the magnetization is the highest, the differences in the internal field distributions are very significant, with the high demagnetizing field orientation having more variance and a tail towards high fields, and the orientation for low demagnetizing field having a tail towards lower fields.

The volume-average of the internal field intensities (marked as the dashed vertical line) is 0.34 T and 0.85 T for the high demagnetizing field and low demagnetizing field orientations, respectively. Figures \ref{fig:6}(b) and \ref{fig:6}(e) are 3D scatter plots, in which each sphere represents the centroid of an element and its colour the internal field intensity at that element. Since the 3D plot only allows us to see the outer layer of the elements, in the figures \ref{fig:6}(c) and \ref{fig:6}(f) the volume-average internal field magnitude across the entire thickness of the sample is plotted in 2D. 

These computations also allow the determination of an effective demagnetizing factor, $D_{eff}$. Since we have access to the magnetization and internal fields in each element, we can relate the average internal field and the applied fields through:

\begin{equation}
    \langle H_{int} \rangle =H_{app}-D_{eff} \langle M(H_{int}) \rangle,
\end{equation}

where rearranging yields the explicit expression:

\begin{equation}
    D_{eff} = \frac{ H_{app}  - \langle H_{int} \rangle}{\langle M(H_{int}) \rangle}.
\end{equation}

In these expressions, the \textit{volumetric} magnetization should be considered, which was obtained by considering the density of Gd$_5$Si$_2$Ge$_2$ from the literature, 7.521 g/cm$^3$ \cite{Gschneidner2005}.

Applying this procedure for the case shown in figure \ref{fig:6}, we obtain $D_{eff}^\parallel$=0.13 and $D_{eff}^\perp$=0.62 for the low demagnetizing field and high demagnetizing field orientations, respectively. These figures vary slightly with temperature and applied field intensity. $D_{eff}^\parallel$ varies between 0.127 and 0.140, and $D_{eff}^\perp$ varies between 0.566 and 0.653. The value of $D_{eff}$ does not convey any information about the shape of the distribution of internal field intensity, but it is useful to quantify the intensity of demagnetization and allow comparisons between different shapes.

These simulations were computed in 0.5 K steps from 260 K to 280 K and 0.2 T steps from 0.2 T to 2.0 T for each applied field orientation, corresponding to a total of 810 computations and allowing us to use equation \ref{eq:1} to compute the $\Delta S_M^{rot}$ between the average internal fields at each configuration (referred in shorthand simply as H1 for the high demagnetizing field orientation and H2 for the low demagnetizing field orientation, which are functions of temperature and applied field amplitude). There was no discernible difference found between computing the average $\Delta S_M$ for the internal field variation of each element or computing $\Delta S_M$ for the overall change in average internal field, so we will proceed with the more intuitive approach of computing the integral between the two average field intensities.

\begin{figure}
    \centering
    \includegraphics[width=160 mm]{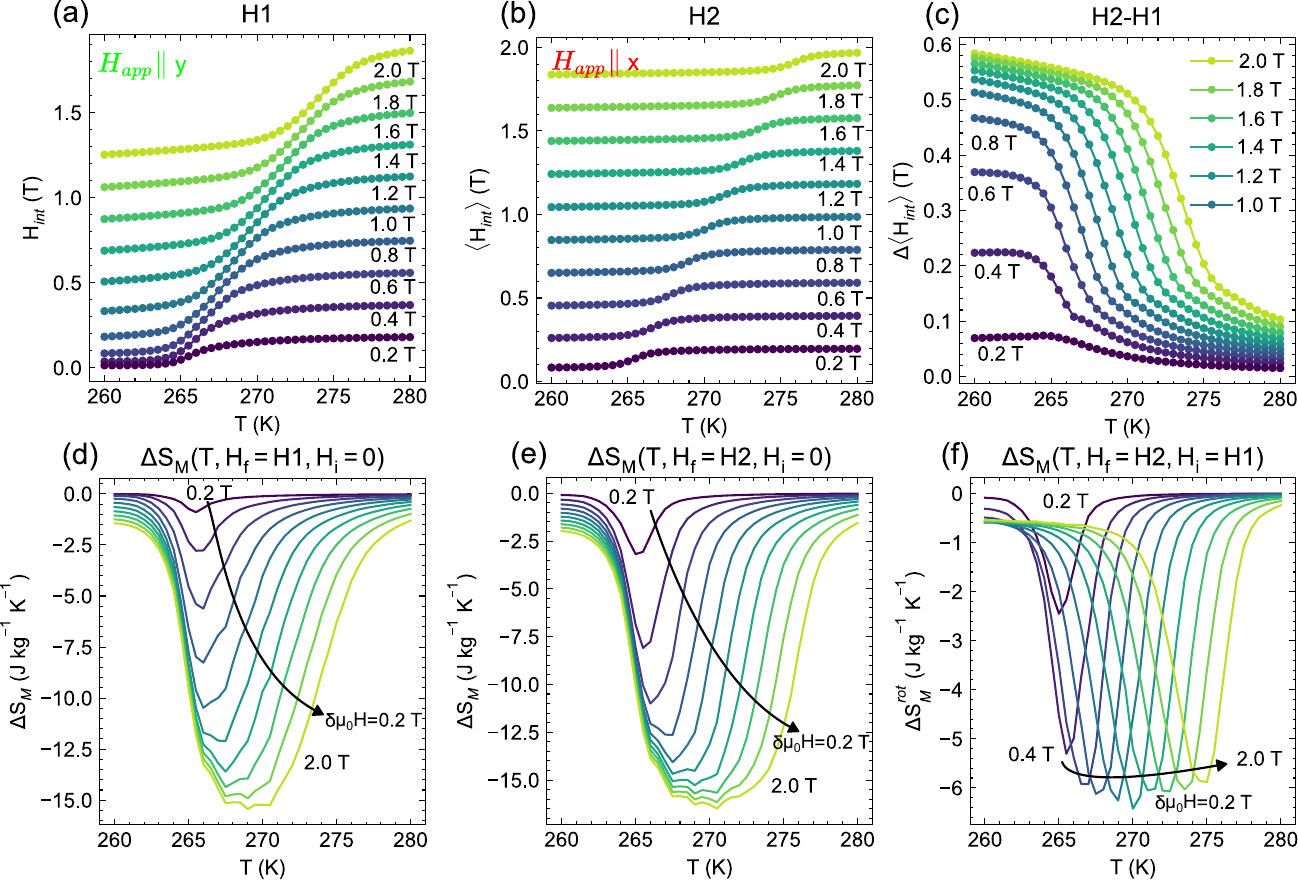}
    \caption{The average internal magnetic field intensities obtained from the \textit{FEMCE} simulations for every temperature and external magnetic field intensity for the (a) high demagnetizing field orientation, (b) low demagnetizing field orientation, and (c) the difference between the two. In the lower part, we see the obtained $\Delta S_M$ for the magnetic field applications along the (d) high demagnetizing field orientation, (e) low demagnetizing field orientation, and (f) the difference between the two, corresponding to $\Delta S_M^{rot}$.}
    \label{fig:7}
\end{figure}

The simulation results are shown in figure \ref{fig:7}. Figure \ref{fig:7}(a) and \ref{fig:7}(b) show the obtained average internal fields for each applied magnetic field orientation. Figure \ref{fig:7}(c) shows the difference between the corresponding average internal fields. For every applied field intensity, a clear reduction of the internal field occurs on cooling at the phase transition temperature, which shifts towards higher temperatures with increasing field intensity. Figures \ref{fig:7}(d) and \ref{fig:7}(e) and correspond to the $\Delta S_M$ of the conventional MCE, which was obtained by computing equation \ref{eq:1} for an initial internal field ($H_i$) intensity of 0 and a final internal field intensity ($H_f$) equal to the corresponding average internal field (H1 or H2, respectively). 

The maximum $\Delta S_M$ obtained for the low demagnetizing field orientation was $\mathrm{-16.49\:J\: K^{-1}\: kg^{-1}}$, in good agreement with a previous study following similar measurement procedures (-18.73 $\mathrm{J\: K^{-1}\: kg^{-1}}$ after correcting for the demagnetizing field) \cite{Bez2018}.

\subsection{Indirect measurements of the rotating magnetocaloric effect - magnetic entropy change during field rotation}

The maximum $\Delta S_M$ obtained for the high demagnetizing field orientation was -15.43 J K$^{-1}$ kg$^{-1}$, corresponding to a reduction of only 6\% with respect to the low demagnetizing field orientation. This could lead one to assume that changing $D_{eff}$ from $\sim 0.6$ to $\sim0.1$ in fact had a very small impact, leading to a small RMCE. However, one may notice from figures \ref{fig:7}(d) and (e) that the widths of the $\Delta S_M$ curves for the two orientations are quite different, in particular at the temperatures above the phase transition. Since the $\Delta S_M$ has a very sharp dependence for lower fields near the phase transition temperature, a change of demagnetizing factor has a small effect on $\Delta S_M$ in that region. At the same time, increasing the applied field intensity leads to a slightly increased peak but also a widening of the $\Delta S_M$ profile towards higher temperatures, which is a typical characteristic of magnetocaloric materials with FOPT \cite{Smith2012}. The impact of the demagnetizing effect is mostly translated through a diminished widening with increased field amplitudes. This contrasts with what has been previously found for Gd \cite{Almeida2024}. The temperature profile of the SOPT of Gd is such that increasing the field raises $\Delta S_M$ in general, instead of a clearly asymmetrical increase.

The entropy change resulting from rotation, $\Delta S_M^{rot}$, corresponds precisely to the difference between the $\Delta S_M$ obtained for each applied magnetic field orientation (which results directly from the fundamental theorem of calculus applied to equation \ref{eq:1}). This is shown in figure \ref{fig:7}(f). Even though the change in average internal field intensity (H2-H1) increases monotonously with applied field intensity, as shown in figure \ref{fig:7}(c), $\Delta S_M^{rot}$ has an extremely non-monotonous field-dependence, in which saturation is very quickly reached while a shift to higher temperature occurs. For instance, the maximum $\Delta S_M^{rot}$ obtained for 0.8 T, -6.12 J K$^{-1}$ kg$^{-1}$, corresponds to 95\% of the global maximum $\Delta S_M^{rot}$ reached, which was -6.42 J K$^{-1}$ kg$^{-1}$ at 1.2 T. These amplitudes of the peak entropy change obtained for the RMCE without changing the applied field intensity (only rotating its orientation) are noteworthy, surpassing the peak values obtained by several other prominent magnetocaloric materials through the conventional MCE even for higher applied field intensities \cite{Law2021}. This further implies that from the perspective of RMCE application, the use of high magnetic fields (>1 T) is unnecessary and, likely, wasteful.

\subsection{Improving the rotating magnetocaloric effect by increasing the aspect ratio and comparison with SOPT gadolinium}

\begin{figure}
    \centering
    \includegraphics[width=140 mm]{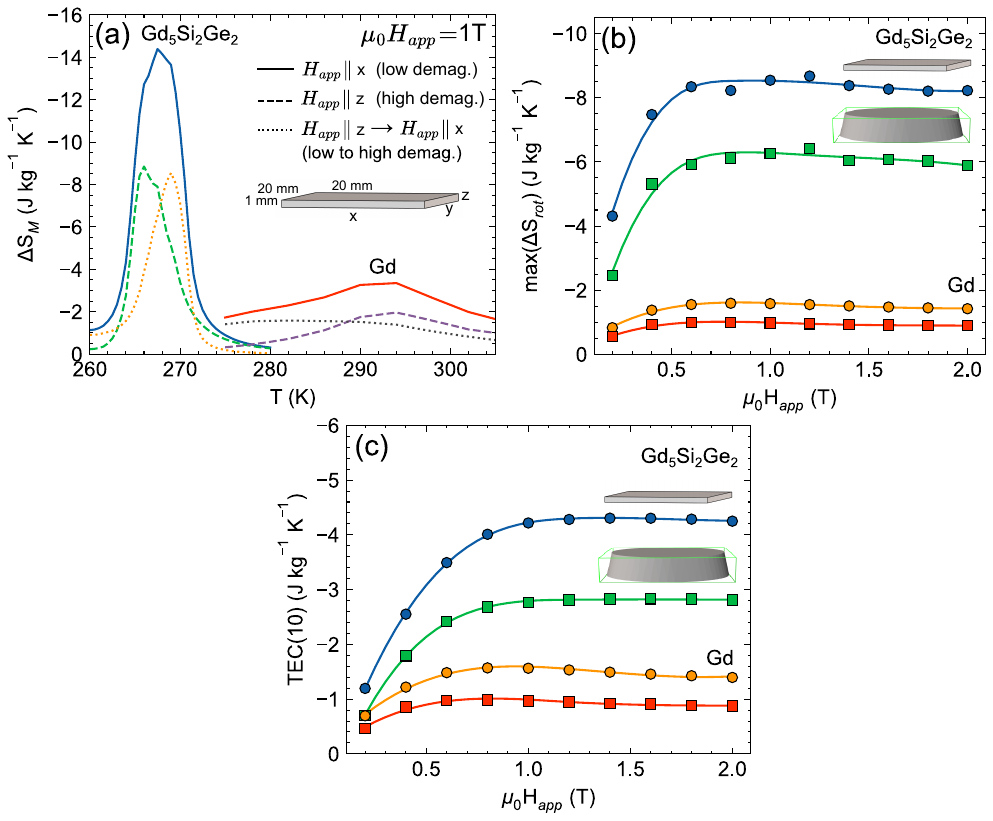}
    \caption{(a) The isothermal entropy change of the conventional (continuous line for low demagnetizing field orientation and dashed line for high demagnetizing field orientation) and rotating (dotted line) MCE of Gd$_5$Si$_2$Ge$_2$ and Gd for a thin plate sample shape with aspect ratio of 20. (b) and (c) show the peak $\Delta S_M^{rot}$ and TEC(10) values as a function of applied magnetic field amplitude for both materials with the thin plate shape (circles) and the experimentally measured shape (squares).}
    \label{fig:8}
\end{figure}

Using a real sample with a large aspect ratio, we already demonstrated significant RMCE values in Gd$_5$Si$_2$Ge$_2$ material. However, as explained in section \ref{sec:Methods}, the RMCE could be further improved in a sample with a more extreme aspect ratio. To explore this possibility, we performed simulations of a thin rectangular prism with an aspect ratio of 20 (20 mm by 20 mm by 1 mm) and compared it to the results of the experimentally measured shape. Furthermore, we perform simulations for a Gd sample with the same thin plate shape, to compare the RMCE of the two materials for the same shape. The results are shown in figure \ref{fig:8}.

Changing the shape from the experimentally measured one to the simulated thin plate decreased $D_{eff}^{\parallel}$ from 0.13 to 0.05 and increased $D_{eff}^{\perp}$ from 0.62 to 0.85. Figure \ref{fig:8}(a) shows the conventional MCE for the low demagnetizing field orientation of the thin plate ("in plane"), the high demagnetizing field orientation ("off plane"), and the RMCE obtained from a rotation between the two. The magnetic entropy change for the thin plate of Gd reached -8.67 J K$^{-1}$ kg$^{-1}$, which is 35\% more than the maximum obtained for the experimentally measured shape. The difference between the $\Delta S_M$ temperature profiles of the RMCE curves obtained for Gd$_5$Si$_2$Ge$_2$ and Gd's SOPT is also very clear, as described above. 

The non-monotonous field-dependence of the RMCE obtained for both materials and shapes is shown in figure \ref{fig:8}(b). The peak RMCE of Gd$_5$Si$_2$Ge$_2$ is over 5 times bigger than that obtained from the simulations for Gd for both shapes. It is clear from this astounding difference that FOPT materials have more potential to achieve high RMCE due to their sharp (in temperature) phase transition, as well as the significant shift of the phase transition temperature with magnetic field.

Finally, it is worth highlighting that the intensity of the magnetic entropy change is inextricably linked to its "width" (in temperature). It is often the case that the high peaks of $\Delta S_M$ displayed by materials with GMCE come at the cost of slimmer profiles in temperature, and the same has been found for the RMCE.  The temperature-averaged entropy change (TEC), previously proposed by L. D. Griffith et al. \cite{Griffith2018}, is a figure of merit that better conveys the temperature profile of $\Delta S_M$. It is defined as

\begin{equation}
    \label{eq:TEC}
    TEC(\Delta T_{lift})= \frac{ \mathrm{max}\left\{ \int_{T_{mid}-\frac{\Delta T_{lift}}{2}}^{T_{mid}+\frac{\Delta T_{lift}}{2}} \Delta S_M (T)dT \right\}}{\Delta T_{lift} },
\end{equation}

where $T_{lift}$ is the temperature lift over which the average is evaluated and $T_{mid}$ is the middle temperature which maximizes the average for the chosen temperature lift.

The field-dependence of the TEC for a temperature lift of 10 K can be seen in figure \ref{fig:8}(c) for both sample shapes. For the chosen temperature lift, Gd$_5$Si$_2$Ge$_2$ still clearly outperforms Gd, however, the wider temperature profile of Gd's $\Delta S_M$ reduces the improvement margin from over 5 times larger to between 2 and 3 times larger for either shape.

\section{Conclusions and outlook}

We have extensively characterized the demagnetizing field-based rotating magnetocaloric effect of a first-order magnetocaloric material, Gd$_5$Si$_2$Ge$_2$, for the first time via two complementary approaches. 

Both the $\Delta T_{ad}^{rot}$ and $\Delta S_{M}^{rot}$ were found to very quickly increase with the applied magnetic field amplitude and then decrease, while shifting toward higher temperatures. The maximum $\Delta T_{ad}^{rot}$ obtained was 1.77 K for 0.8 T, and the maximum $\Delta S_{M}^{rot}$ was -6.42 J K$^{-1}$ kg$^{-1}$, for 1.2 T. For 0.8 T, $\Delta S_{M}^{rot}$ already reaches -6.12 J K$^{-1}$ kg$^{-1}$, 95 \% of its maximum value, higher than the maximum $\Delta S_{M}$ obtainable in the conventional MCE of Gd with 1.0 T. Our study suggests that the RMCE is fundamentally suited for low magnetic field applications, particularly in first-order magnetocaloric materials.

Finally, the already impressive peak $\Delta S_M^{rot}$ of the experimentally measured shape was found to increase by 35\% in a sample with a more extreme aspect ratio of 20. A proportional increase of the maximum $\Delta T_{ad}^{rot}$ would raise it over 2 K, which recently has been suggested as an important threshold required for developing efficient refrigeration devices \cite{Klinar2024}, reinstating the demagnetizing field-based RMCE's relevance in magnetic refrigeration, in particular for magnetocaloric materials with FOPT. While the non-monotonous field-dependence of the RMCE had been previously found for Gd, the overall temperature-dependence of the RMCE in Gd$_5$Si$_2$Ge$_2$ was fundamentally different due to its first-order phase transition, which is much sharper in temperature for all magnetic fields measured and significantly shifts towards high temperatures with increasing applied field. 

The results of our work clearly show the significant influence of the material's shape both for the conventional and rotating MCE, hopefully informing future studies of GMCE materials.

Regarding heat pumping applications, we underline that, while the RMCE enables a different kind of device design, where no field intensity modulation is needed, potentially reducing the total volume of permanent magnets required, some new challenges are also expected. Namely, if the magnetic field region used in the device is cylindrical, packing as much magnetocaloric material as possible within it (porosities are usually below 50\% \cite{Kitanovski2015}) will inevitably lead to an overall reduction of the aspect ratio, thus decreasing the RMCE in comparison with what is possible with an isolated, high aspect ratio plate \cite{Christensen2011}. This leads to a complex optimization problem, where increasing material content simultaneously decreases the intensity of the effect, while also altering the thermofluidic properties. This investigation will be detailed in future works.

\section*{Acknowledgements}

The authors acknowledge FCT and its support through the projects LA/P/0095/2020, UIDB/04968/2025, and UIDP/04968/2025. R. Almeida thanks FCT for his PhD grant with reference 2022.13354.BD and DOI \\ https://doi.org/10.54499/2022.13354.BD. This project has received funding from the European Union’s Horizon Europe research and innovation programme through the European Innovation Council under the grant agreement No. 101161135 – MAGCCINE. This work was also developed within the scope of the project CICECO-Aveiro Institute of Materials, UIDB/50011/2020, UIDP/50011/2020 \& LA/P/0006/2020, financed by national funds through the FCT/MCTES (PIDDAC). The work at the Ames National Laboratory (Y.M. and K. D.-A.) on material's synthesis and manuscript editing was supported by the Division of Materials Science and Engineering of the Office of Basic Energy Sciences, Office of Science of the U.S. Department of Energy (DOE). Ames National Laboratory is operated for the U.S. DOE by Iowa State University of Science and Technology under Contract No. DE-AC02-07CH11358.

\appendix

\section{Magnetic and structural characterization}

\begin{figure}
    \centering
    \includegraphics[width=75 mm]{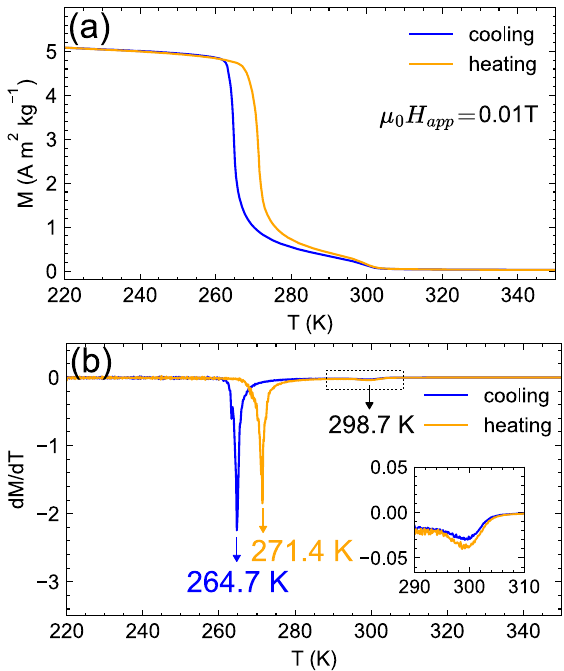}
    \caption{(a) Isofield magnetization on cooling and heating with a 0.01 T applied magnetic field, measured at a temperature change rate of 2 K/min. (b) The derivatives of the two curves, showing peaks at the phase transition temperatures.}
    \label{fig:A1}
\end{figure}

The main phase of Gd$_5$Si$_2$Ge$_2$ exhibits the first-order magnetostructural phase transition behind its GMCE near 270 K. From low to high temperature, it transforms from an orthorhombic-II ferromagnetic phase to a monoclinic paramagnetic phase \cite{MELIKHOV2015143}. At higher temperatures, a small fraction of Gd$_5$Si$_2$Ge$_2$ phase remains orthorhombic and ferromagnetic, exhibiting a second-order phase transition to a paramagnetic state near 300 K. Our findings, shown in figure \ref{fig:A1}, are consistent with the literature.

\begin{figure}
    \centering
    \includegraphics[width=90 mm]{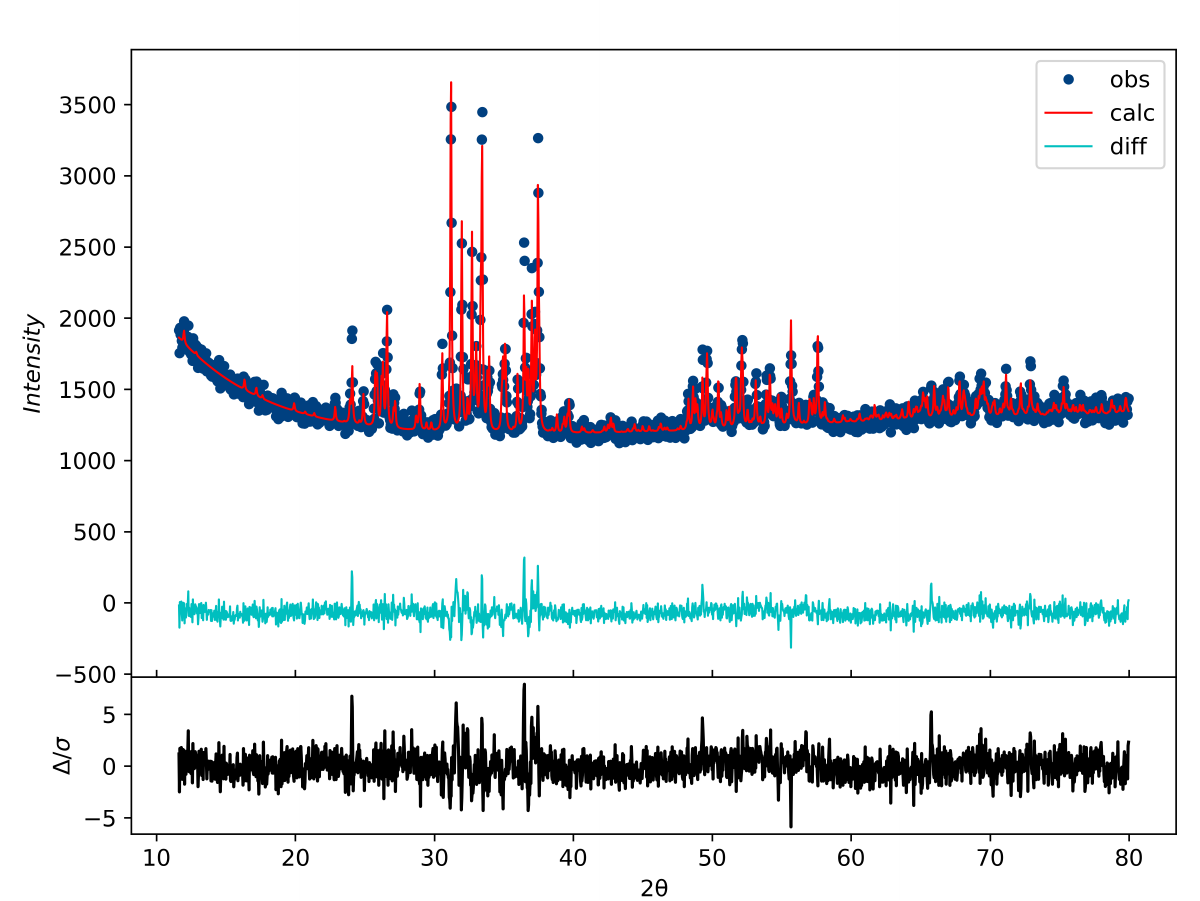}
    \caption{Powder X-ray diffraction measurements of the prepared Gd$_5$Si$_2$Ge$_2$ at room temperature and Rietveld refinement results.}
    \label{fig:A2}
\end{figure}

Another sample was extracted from the same batch as the other samples used in this study and measured in powder X-ray diffraction at room temperature. The Rietveld refinement results are shown in figure \ref{fig:A2}. The pattern show a clear match (GOF = 1.3) with the reported monoclinic Gd$_5$Si$_2$Ge$_2$ crystal structure with  lattice parameters: $a = 0.75788$ nm, $b = 1.48011$ nm, $c= 0.77804\:\mathrm{nm}$, and $\gamma$ = 93.17 deg \cite{Mudryk2005}.

\printcredits

\bibliographystyle{ieeetr}

\bibliography{bib}



\end{document}